\documentclass[aps, prb, superscriptaddress,reprint,citeautoscript]{revtex4-2}
\usepackage{graphicx}
\usepackage{dcolumn}
\usepackage{bm}
\usepackage{amsmath}

\usepackage{color}
\usepackage[dvipsnames]{xcolor}
\usepackage{bm,graphicx,hyperref}
\newcommand{\be}{\begin{equation}}
\newcommand{\ee}{\end{equation}}

\begin{document}
	
\date{\today}
\title{Collective Excitations in 2D Materials}
\author{Aleksandr Rodin}
\affiliation{Centre for Advanced 2D Materials, National University of Singapore, 117546}
\affiliation{Yale-NUS College, 16 College Avenue West, 138527, Singapore}
\author{Maxim Trushin}
\affiliation{Centre for Advanced 2D Materials, National University of Singapore, 117546}
\author{Alexandra Carvalho}
\affiliation{Centre for Advanced 2D Materials, National University of Singapore, 117546}
\author{and A. H. Castro Neto}
\affiliation{Centre for Advanced 2D Materials, National University of Singapore, 117546}
\affiliation{Department of Materials Science Engineering, National University of Singapore, 117575}

\begin{abstract}
Research on 2D materials has been one of the fastest-growing fields in condensed
matter physics and materials science in the past 10 years. The low dimensionality and strong
correlations of 2D systems give rise to electronic and structural properties, in the form of
collective excitations, that do not have counterparts in ordinary 3D materials used in modern
technology. These 2D materials present extraordinary opportunities for new technologies, such
as in flexible electronics. In this Review, we focus on plasmons, excitons, phonons and magnons in
2D materials. We discuss the theoretical formalism of these collective excitations and elucidate
how they differ from their 3D counterparts.
\end{abstract}    

\maketitle

\section{Introduction}
\label{sec:Introduction}

As a rule, the basis for the understanding of properties of materials relies on their non-interacting, or single particle, properties. The ``particle in a box" model for 3D metals, proposed by Sommerfeld \citep{ashcroft2011solid} at the beginning of the 20$^{\it th}$ century, is an example of how an oversimplified model of non-interacting electrons can provide explanations for physical behaviour and make predictions that have been confirmed experimentally innumerable times. It took many decades of study, and serious theoretical work, culminating in the Landau theory of the Fermi liquid \citep{nozieres1963}, to understand why and how charged particles that interact via strong long-range repulsive forces could behave as if they did not ``feel" each other. At the same time, Bloch's theorem and the concept of band structure of crystals \citep{ashcroft2011solid} for non-interacting electrons have been at the core of virtually all modern technology based on semiconductors.   

Driven by either pure scientific curiosity or by overwhelming technological demands, the study of new materials has been revolutionary, not evolutionary, in the last decades. The isolation of graphene in 2004 \citep{geimnovoselov2004} and the emergence of a new class of 2D materials that became easily accessible experimentally has driven intense research in the last 16 years. It became evident in the last decade that, although several concepts in 3D solid-state physics still apply to 2D materials, the standard model of the theory of electrons in solids has to be revised to explain the unique properties of 2D materials \citep{review-graphene}. We note that, when referring to 2D materials, we mean crystalline systems with one of their three dimensions being of atomic thickness, as opposed to quasi-2D systems, where the electrons are confined to a plane within a three-dimensional material.

\subsection{Electron interaction}

The concepts of screening and non-interacting quasiparticles lie at the heart of the theory of 3D metals. In vacuum, two charged particles, such as electrons, interact with each other via the Coulomb potential that decays as {\it e}$^2$/{\it r}, where {\it e} is the charge of the electron and {\it r} is the distance between the particles. The electric field generated by these charges permeates the whole space around them and creates a substantial repulsive force that keeps the charges apart. However, paraphrasing Anderson \citep{anderson1972}, ``more is different", namely, when $10^{23}$ electrons in a material are forced to share the space of a 3D crystal, they can move around to shield, or screen, the strong 3D electric field. The final result of this electronic ``dance" is a significantly reduced interaction $\exp\{-{\it r}/\ell\}$/{\it r} where $\ell$ is the so-called screening length that depends on the electronic density in the material \citep{mahan2000many}. The new character that emerges from this scenario, the quasiparticle, can be seen as an electron ``dressed" by the cloud of all other electrons \citep{nozieres1963}. Instead of a long-range interaction, the quasiparticles interact through weak short-range forces, validating the non-interacting model proposed by Sommerfeld. It is essential to notice that this result comes about because electrons, and the electric field they carry with them, propagate in 3D.

In a 2D material, electrons are confined by the attractive potential of the lattice ions (with a positive charge) that halts their motion perpendicular to the plane of the material. Nevertheless, the electric field produced by the electrons still propagates in the 3D space. Therefore, the electric field cannot be fully screened because there are no free charges that can move around to shield the strong interaction between electrons. This lack of screening is one of the main characteristics of 2D materials \citep{Ando1982,kotov2012}. 

Hence, 2D materials are strongly correlated by nature, and the concept of quasiparticles and single-particle excitations is debatable. Furthermore, collective electron modes in 2D metals such as plasmons become dispersive, that is, their energy becomes momentum-dependent, and their speed of propagation (or group velocity) becomes divergent at long wavelengths and low energies, leading to serious implications from the experimental point of view. 2D plasmons cannot be observed experimentally in the same way of their 3D counterparts, and new experimental techniques must be used.

In 2D semiconductors, this state of affairs is even more dramatic. In the 3D semiconductors used in modern-day technology, an electron in the conduction band and a hole in the valence band can bind in the form of Hydrogen-like atom called an exciton. These 3D excitons are very well described by the hydrogen model (H-model) that is one of the most significant theoretical achievements of quantum mechanics. In 2D, the situation is rather different. The lack of screening discussed previously invalidates many of the assumptions of the H-model. Firstly, in 3D, only states close to the conduction and valence band edges are sufficient to describe the physics of exciton formation. This so-called ``effective mass" approximation essentially retraces the Sommerfeld model of metals of non-interacting particles. In 2D semiconductors, the strong interactions require most of the states in the band structure for a precise description of the excitonic state and its binding energy. Excitons can be observed experimentally in photoluminescence experiments, but cannot be calculated reliably from the H-model. Furthermore, because of the same strong interactions, a plethora of other excitations such as bi-excitons (bound state of two electron-hole pairs), trions (bound states of two electrons and one hole), quintons (bound states of three electrons and two holes) and so on, start to play an essential role in the interpretation of the experimental data.

\subsection{Structural properties}

Another essential characteristic of 2D materials is their inherent softness in the sense that it costs much less energy to create collective states. Consider, for instance, collective lattice vibrations, or phonons. In 3D, an atom has its motion constrained, by the presence of neighbouring atom in all three directions {\it (x,y,z)}. In 2D materials one of the directions of motion, say {\it z}, is unconstrained leading to new phonon modes, ``drum-head" modes, whose energy, $E$, depends on the wavelength, $\lambda$, as $E \propto 1/\lambda^2$, becoming vanishingly small in energy at very long wavelengths. In thermal equilibrium, at a temperature $T$, the number of such modes becomes divergent leading to structural instabilities such as crumpling \citep{review-graphene}, which is the basis of the Hohenberg-Mermin-Wagner theorem that states the absence of long-range order in 2D \citep{mermin1966absence}. The same type of argument applies to 2D ferromagnetic systems where the collective magnetic states, or magnons, have the same dispersion relation. Therefore, 2D magnetic systems are prone to different types of magnetic collective states such as skyrmions, merons, etc., increasing even further the zoo of collective states that live in 2D. 

In this Review, we discuss various states in turn. We begin by addressing plasmons in 2D metals. Following this, we focus on excitonic states in 2D semiconductors.
Next, we provide an overview of phonons and collective states in 2D magnets. We conclude with a perspective on future directions of the field.

\section{Plasmons}
\label{sec:Plasmons}

Oscillations in electronic charge density give rise to plasmons. Even though, just like phonons, plasmonic modes can exist in systems of any dimensionality, 2D plasmons possess a set of characteristics that distinguish them from their 3D counterparts.

Subjecting an electronic system, regardless of its dimensionality, to an external perturbing potential $\phi_\mathrm{ext}\left(\mathbf{r},t\right)$ redistributes the charges and creates an induced electronic density $\rho_\mathrm{ind}\left(\mathbf{r},t\right)$. This induced density, in turn, generates its own potential so that the total potential becomes
\begin{equation}
    \phi_\mathrm{tot}\left(\mathbf{r},t\right) = \phi_\mathrm{ext}\left(\mathbf{r},t\right) 
    + 
    \int d\mathbf{r}'\rho_\mathrm{ind}\left(\mathbf{r}',t\right)V_{\mathbf{r} - \mathbf{r}'}\,,
    \label{eqn:phi_tot}
\end{equation}
where $V_{\mathbf{r}}$ is the Coulomb term, $\mathbf{r}$ is the coordinate at which we want to know the total potential, and $\mathbf{r}'$ is the coordinate of the source density $\rho_\mathrm{ind}\left(\mathbf{r}',t\right)$. It is possible to relate $\rho_\mathrm{ind}\left(\mathbf{r},t\right)$ to $\phi_\mathrm{tot}\left(\mathbf{r},t\right)$ through the polarisation function $\Pi\left(\mathbf{r}, t\right)$~\citep{ashcroft2011solid, mahan2000many, Bruus2002}
\begin{equation}
    \rho_\mathrm{ind}\left(\mathbf{r},t\right) = \int d\mathbf{r}' \int dt' \phi_\mathrm{tot}\left(\mathbf{r}',t'\right)\Pi\left(\mathbf{r} - \mathbf{r}',\,t - t'\right)\,.
    \label{eqn:rho_ind}
\end{equation}
In the Fourier space, the convolutions in Eqs.~\eqref{eqn:phi_tot} and \eqref{eqn:rho_ind} are replaced by products so that
\begin{equation}
    \phi_\mathrm{tot}\left(\mathbf{q},\omega\right) = \phi_\mathrm{ext}\left(\mathbf{q},\omega\right) 
    + 
   \Pi\left(\mathbf{q},\omega\right)V_{\mathbf{q}}\phi_\mathrm{tot}\left(\mathbf{q},\omega\right)\,,
    \label{eqn:phi_tot_FT}
\end{equation}
where $\mathbf{q}$ is the wave vector and $\omega$ is the frequency of oscillations. The plasmonic modes are obtained from Eq.~\eqref{eqn:phi_tot_FT} by requiring that self-sustaining oscillations in $\phi_\mathrm{tot}\left(\mathbf{q},\omega\right)$ exist in the absence of the external perturbation. In other words, $\Pi\left(\mathbf{q},\omega\right)V_{\mathbf{q}} = 1$, which gives the relationship between the frequency of the oscillations $\omega$ and the corresponding wave vector $\mathbf{q}$.

The polarization function can be computed from the Lindhard formula~\citep{ashcroft2011solid, mahan2000many}. For a single-band parabolic dispersion $\varepsilon_\mathbf{q}$, the small-$\mathbf{q}$ polarization is $\Pi_\mathrm{2D}\left(\mathbf{q},\omega\right) = \mu q^2 / (\pi \hbar^2\omega^2)$ and $\Pi_\mathrm{3D}\left(\mathbf{q},\omega\right) = 2n_0\varepsilon_\mathbf{q}/(\hbar^2\omega^2)$, where $\mu$ is the chemical potential, $n_0 = k_F^3 / 3\pi^2$ the electronic density ($k_F$ is the Fermi momentum), $\varepsilon_\mathbf{q} = \hbar^2 q^2/(2 m^*)$ is the free electron dispersion with effective mass $m^*$. By comparing the two- and three-dimensional polarization functions, one can see that they both are proportional to $q^2/\omega^2$. Even though these results were obtained for a massive isotropic dispersion, they also hold for Dirac~\citep{Wunsch2006, Hwang2007} and anisotropic bands~\citep{Low2014, Rodin2015}.

The key difference between the dimensionalities is in the Coulomb term: $V_\mathbf{q}^\mathrm{2D}\propto q^{-1}$, while $V_\mathbf{q}^\mathrm{3D}\propto q^{-2}$. From this, $\Pi\left(\mathbf{q},\omega\right)V_{\mathbf{q}} = 1$ states that, in 3D, the plasma frequency $\omega_p$ is a momentum-independent quantity. For real 3D materials, the magnitude of $\hbar\omega_p$ is of the order of electron volts, making plasmons high-energy, weakly-dispersive excitations. Although $\omega_p$ does depend on the carrier density, this quantity cannot be changed appreciably in a 3D sample so that $\omega_p$ cannot be tuned experimentally. This large $\omega_p$ value means that substantial energy input is necessary to excite the charge oscillations in 3D. As a consequence, one of the ways that the plasmons in 3D are confirmed experimentally is by sending energetic electrons through thin metallic films to observe that the energy lost by the electrons occurs in multiples of $\hbar\omega_p$.~\citep{Marton1962}

The situation in 2D is very different with $\omega_p\propto\sqrt{q}$,~\citep{Ando1982} meaning that the energy of the excitation $\hbar\omega$ can be made arbitrarily small by increasing the wavelength of the charge oscillations, resulting in the so-called ``soft modes". A soft mode means that the energy of the excitation vanishes in the long-wavelength limit. This energy dependence is radically different from the 3D plasmon, where, even as $\mathbf{q} \rightarrow 0$, $\hbar\omega_p$ remains finite. Also, unlike the 3D case, it is possible to change the charge density in a sample by gating, allowing one to tune the wavelength of the plasmonic modes.

Another important consequence of the $\sqrt{q}$ dispersion has to do with the plasmon-electron interaction. Because the speed of the plasmons varies with their momentum, it is possible to match this speed with the electronic Fermi velocity. This speed matching leads to a strong coupling between electrons and the collective modes, giving rise to plasmarons,~\citep{Polini2008} which have been observed experimentally.~\cite{Bostwick2010}

In a many-body electronic system, individual particle states can be labelled by their momentum $\mathbf{k}$. The polarisation function describes a process of transferring a particle from a filled state to an empty one. Crucially, the state in which the system ends up after such a transfer is not an eigenstate of the many-body Hamiltonian, being, instead, a superposition of many eigenstates. Because of this, the freshly-moved particle will not remain in its new state $\mathbf{k}' $ as the individual many-body states evolve in time. Another way to describe what happens is that the transferred particle scatters off other electrons. Lindhard polarisation function neglects this effect and assumes that the transferred particle stays in its new state. While this might appear to be a serious shortcoming, we will see below that it works well for the systems of interest.

In addition to their tunability, 2D plasmons have the benefit of being observable in real space. To excite these modes, one needs to provide an external perturbation to a 2D system of the correct $\omega$ and $\mathbf{q}$, determined by the solution of $\Pi_\mathrm{2D}\left(\mathbf{q},\omega\right) V_\mathbf{q} = 1$. It would appear reasonable that illuminating the system with the light of the desired frequency would excite the specific plasmon mode. The issue with this approach is that the wavelength of light with frequency $\omega$ is different from the wavelength of the plasma oscillations at the same $\omega$. Therefore, simply shining the light on the sample will not create plasmons because the momenta do not match.

To solve the momentum-mismatch problem, experimental groups use scanning near-field optical microscopy (SNOM).~\citep{Hillenbrand2000, Fei2011, Zhang2012, Fei2012, Chen2012} In this technique, the electronic system is scanned by the tip of the atomic force microscope (AFM) which is being illuminated by a light beam of the desired frequency. The polarisable tip scatters the incident light and creates evanescent waves covering a range of different momenta, peaked at the inverse of the tip curvature radius, decoupling wavelength from frequency. A portion of these evanescent waves carries the ``correct" momentum to excite the plasmons with frequency $\omega$.

The plasmons launched by the tip propagate radially and reflect off defects and edges. Continuously illuminating the system produces a non-equilibrium steady state with standing waves, and the tip detects the plasmonic signal at its location. If the reflected wave is in phase with the tip polarisation, we get a constructive interference at the tip position and observe an enhanced response. Similarly, if the reflected wave is out of phase with the tip, the signal will be diminished.

\begin{figure*}
    \includegraphics[width = 6in]{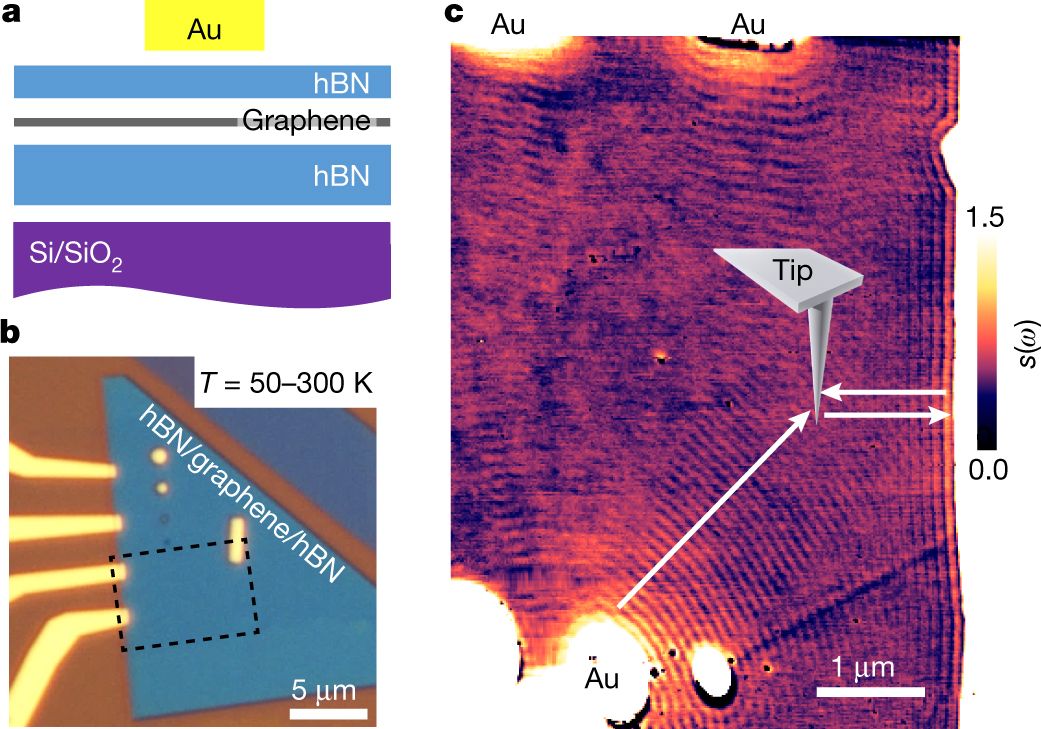}
    \caption{A sketch showing the cross section of the device (a) and an optical image (b) of a graphene sheet encapsulated by hBN. SNOM data collected at 50K is shown in panel (c). The plasmonic modes are launched by the tip and the gold islands. Reproduced with permission from Ref.~\citep{Ni2018}, Springer Nature Limited.}
    \label{fig:SNOM}
\end{figure*}

An experimental set-up used for plasmonic imaging is shown in Fig.~\ref{fig:SNOM}. In Fig.~\ref{fig:SNOM}(a), one can see a schematic illustration of the device with the plasmon-hosting graphene encapsulated by hexagonal boron nitride (hBN), situated on top of the Si/$\mathrm{SiO}_2$ gate. Gold microstructures are deposited on top of the sample. Figure~\ref{fig:SNOM} (b) shows an optical image of the sample.

By moving the illuminated AFM tip around, one can record the response throughout the sample to obtain images such as those shown in Fig.~\ref{fig:SNOM}(c). To reiterate: the data do not represent instantaneous snapshots. Instead, each point is a total signal arising from the tip perturbation plus the contribution due to reflected waves, explaining why the signal is essentially invariant as the tip moves along the edge of the sample.

Close to the edge, one can observe a series of fringes due to alternating constructive/destructive interference. As expected in an interference problem, the spacing between the fringes is one-half of the plasmonic wavelength, determined by the chemical potential and the light frequency. Figure~\ref{fig:SNOM}(c) also demonstrates the ability of plasmons to ``squeeze" the electromagnetic field. The incident wave generated by an $886~\mathrm{cm}^{-1}$ laser has the wavelength of approximately $11.3$~$\mu$m. The plasmon wavelength, on the other hand, is close to 200~nm while retaining the same frequency.

In the earliest work of real-space plasmon imaging~\citep{Fei2012, Chen2012}, it was observed that the amplitude of the signal oscillations gets weaker as the distance from the reflector (edge or defect) increases. Although one does expect this as a consequence of the circular wave spreading out, the experimentally-observed decay is exponential rather than power-law, raising the question of what determines the plasmon lifetime.

Since plasmonic modes are composed of a periodically modulated electronic density, the decay of the modes implies that the constituent electrons stop moving in unison. The reason why an electron would change its trajectory is scattering, and, in the system of interest, there are three main scattering pathways: electron-electron, electron-phonon, and electron-disorder. To understand which of these processes dominates the plasmonic decay, one should assess them individually. The situation with the electron-electron interaction is delicate, and we will discuss it shortly, after addressing the other two mechanisms.

To minimise the effects of the disorder, one can use ultraclean encapsulated samples. Lowering the temperature can also be used to freeze out the lattice vibrations. Both these steps were taken in Ref.~\citep{Ni2018}, see Fig.~\ref{fig:SNOM}. In BN-encapsulated graphene at cryogenic temperatures, plasmon oscillations retain their shape over multi-micrometre distances as the propagation length increases from a few wavelengths to fifty. Increasing the temperature suppresses the plasmon propagation, indicating that, in clean samples, it is the electron-phonon scattering that plays the dominant role in plasmon decay at these frequencies.

The fact that electron-phonon interaction plays the leading role in determining the plasmon lifetime does not mean that electron-electron interaction is inconsequential. After all, it is the Coulomb repulsion that provides the restoring force for the charge oscillations. As is often the case in oscillating systems, it is all about the time scales. Electrons in a metal have a characteristic collision frequency which determines the time that it takes for a perturbation in a system to dissipate. If the perturbation is applied substantially more quickly than the electrons can equilibrate, one can neglect the equilibration process involving individual electrons and only consider the density-density interaction, which is at the heart of the Lindhard formalism. In other words, the external potential creates plasmons faster than they can decay via electron-electron scattering. As the frequency of the external perturbation is lowered, electron-electron interaction becomes relevant, and plasmons experience strong damping. To include this effect, we have to return to the polarisation function and go beyond the Lindhard approximation.

\section{Excitons}
\label{sec:Excitons}

A layered semiconductor can be thinned down to a monolayer, becoming, essentially, a 2D material, where charge carrier motion (if any) is heavily restricted along the
out-of-plane direction. The electrostatic field, however, remains 3D because the field lines can extend beyond the semiconducting layer, where the dielectric 
screening is limited. The resulting electrostatic interactions are therefore much stronger in 2D layers than in the 3D samples made of the same semiconducting material.

\begin{figure}
  \includegraphics[width=\columnwidth]{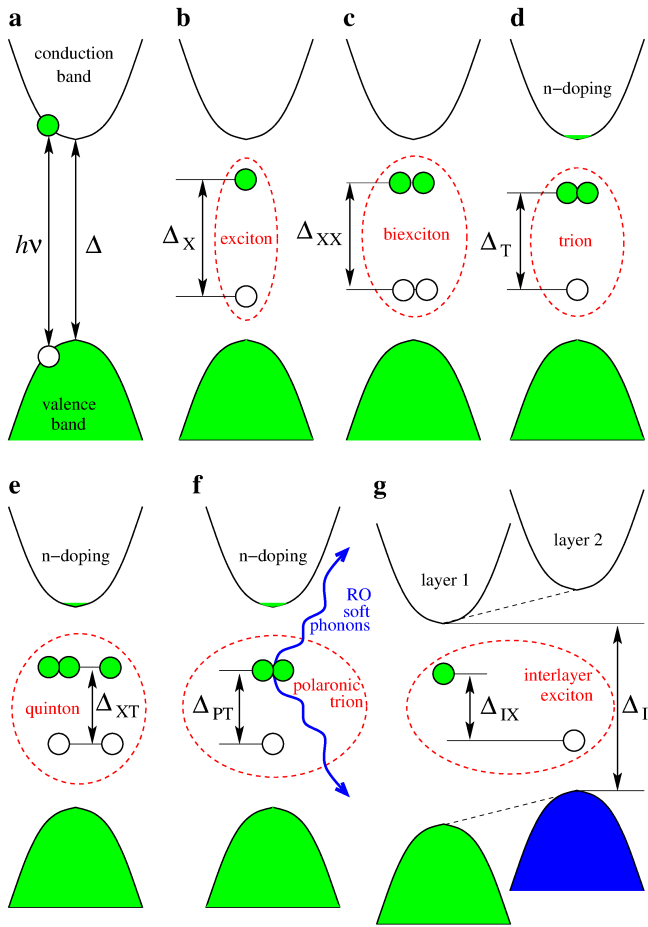}
  \caption{Optically excited quasiparticles with the photoluminescence emission
energies ($\Delta_X$, $\Delta_T$, $\Delta_{XX}$, $\Delta_{XT}$, $\Delta_{PT}$, $\Delta_{IX}$) indicated for each quasiparticle.
{\bf a} Free electron-hole pairs can be created in a direct-gap semiconductor by absorbing a photon of energy $h\nu$.
{\bf b} Excitons (electron-hole bound states) occur as a result of strong Coulomb attraction.
{\bf c} Trions (exciton-electron bound states) form owing to dipole-electron interactions in the presence of excess electrons.
{\bf d} Biexcitons (exciton-exciton bound states) occur due to dipole-dipole interactions.
{\bf e} Quintons may occur due to coupling between trions and excitons.
{\bf f} Polaronic trions may form due to Fr\"ohlich interactions dressing bare trions with the soft RO phonons provided by some special substrates.
{\bf g} Interlayer excitons may form in double-layer 2D semiconductor structures.}
  \label{fig1}
\end{figure}

Optically excited electrons and holes, having opposite charge, experience Coulomb attraction, see Fig. \ref{fig1}. An electron-hole pair forming a bound state is known as {\em exciton}. Excitonic features can be observed in {\em e.g.} absorbance, reflectance, photocurrent, and photoluminescence (PL) spectra~\cite{wang2018colloquium,mak2016photonics,sun2017review}.
In 2D semiconductors, however, the interactions are so strong that it is possible to resolve additional spectral features corresponding to much more complex electron-hole formations.

The physics of optically excited electron–hole complexes (which may comprise any number of electrons and holes) can be understood by using the same 
quasiparticle concept as for excitons. Each quasiparticle species can be characterized by properties such as electric charge, mass and binding energy.
We start from the theoretically simplest case, the excitonic Hamiltonian given by
\begin{equation}
H_X=\frac{\mathbf{P}_X^2}{2m_X} + \frac{\mathbf{p}_X^2}{2\mu_X} 
+ V_X\left(\mathbf{r}_X\right),
\label{H_X}
\end{equation}
where $\mathbf{P}_X=-i\hbar \partial/\partial \mathbf{R}_X$, 
$\mathbf{p}_X=-i\hbar \partial/\partial \mathbf{r}_X$ are momenta operators for the center-of-mass and relative electron-hole motion, respectively, and $V_X\left(\mathbf{r}_X\right)$ is the electron-hole potential. The relative and center-of-mass coordinates are defined in a 2D plane as
$\mathbf{r}_X=\mathbf{r}_e - \mathbf{r}_h$,
$\mathbf{R}_X=(m_e \mathbf{r}_e + m_h \mathbf{r}_h)/m_X$,
where $m_{e,h}$ and $\mathbf{r}_{e,h}$ are the electron (hole) masses and coordinates, respectively. 
The former determine the total mass $m_X=m_e+m_h$ as well as the reduced mass $\mu_X=m_e m_h/m_X$.

The simplest excitonic model is based on a hydrogen-like solution
of the Schr\"odinger equation with the Coulomb potential 
$V_X\left(\mathbf{r}_X\right)=-e^2/(\bar\epsilon r_X)$,
where $\bar\epsilon$ is the average relative dielectric permittivity of a semiconductor.
The model works reasonably well in 3D semiconductors; however, it may fail in the 2D limit for several reasons.

The most important reason is known for decades 
\cite{keldysh1997excitons,keldysh1979coulombENG,keldysh1979coulombRUS,rytova1965coulomb}
and has been revised recently
\cite{cudazzo2011dielectric,berkelbach2013theory,PRB2015latini,kylanpaa2015binding,myPRB2019}:
once the thickness of a semiconductor layer becomes comparable with or smaller than the exciton size, the field lines connecting an electron and a hole extend beyond the semiconducting material that reduces effective screening. Solving the Poisson equation for a 2D semiconductor with the in-plane dielectric polarizability $\chi$ and average dielectric
permittivity of the environment $\bar\epsilon$ one can find
the 2D Fourier transform of the electron-hole potential given by
$V_X(\mathbf{q})=-2\pi e^2/(\epsilon_q q)$, where 
$\epsilon_q=\bar\epsilon(1+r_s q)$, and $r_s=2\pi\chi/\bar\epsilon$.
The real-space potential can be expressed in terms of the Bessel and Struve functions. 2D excitons can be described by a hydrogen-like model
as long as the excitonic ground-state radius is substantially larger than the screening length $r_s$, so that
the quasiclassical radius falls onto the Coulomb-like tail of the potential.
In the opposite limit, the electron-hole potential
does not follow Coulombic $-1/r$ functional dependence, but
turns out to be a much weaker function of the form $\ln(r/r_s)$.
The binding energy is also a much weaker function of $\bar\epsilon$
in this case, as compared with its hydrogenic version.

The less obvious reasons for the hydrogen-like model failure lie in limitations imposed by the effective mass approximation.
First of all, the reduced electrostatic screening in the 2D limit makes the real-space exciton radius smaller
but, at the same time, the exciton gets larger in the momentum space involving the states with
higher momenta beyond the regions of parabolic band dispersion \cite{myPRB2019}.
This can easily happen for the valence band,
which may have dispersion peculiarities not that far from its top
in thin semiconducting layers \cite{zolyomi2013Ga2Se2,rybkovskiy2014Ga2Se2}.
Hence, the tightly-bound excitons may not always be described in terms of the electron and hole effective masses
despite the real-space exciton size remaining larger than the lattice constant justifying the Wannier-Mott picture.
Besides the non-parabolicity of the bands, the effective mass approximation does not take into account
the Berry phase that can influence quantisation of the relative electron-hole motion changing the exciton spectrum
\cite{PRL2015zhou,PRL2015srivastava,PRL2018trushin,PRB2016trushin,EPL2014goerbig}.

Hence, the excitons in 2D semiconductors are in general {\em non-hydrogenic}; however, the hydrogen-like model can still be used to understand the excitonic ground state behaviour qualitatively. In this case, the Schr\"odinger equation can be solved explicitly, and the exciton binding energy reads
$E_X=2\mu_X e^4/(\bar\epsilon^2 \hbar^2)$.
The binding energy $E_X$ is of the order of 100 meV in 2D semiconductors
(e.g. $0.32$ eV in WS$_2$ \cite{WS2-chernikov}, $0.37$ eV in WSe$_2$ \cite{WSe2-he}) when deposited on a SiO$_2$ substrate.
Up to $\sim 50$\%  modulation of binding energy can be achieved by changing the dielectric substrate \cite{lin2014dielectric,borghardt2017engineering,gupta2017direct,rodin2014excitons}.
Even greater modulation  can be reached by making use of free-carrier screening in nearby graphene \cite{raja2017coulomb,qiu2019giant}.

The screening in 2D semiconductors is so weak that even dipole-electron and dipole-dipole interactions turn out to be strong enough to form bound states
between one exciton and one electron, resulting in a trion, between two excitons to form a biexciton, or even between one trion and one exciton to create a quinton.
 In what follows, we review the quasiparticle zoo on an equal theoretical footing.

\begin{figure}
  \includegraphics[width=\columnwidth]{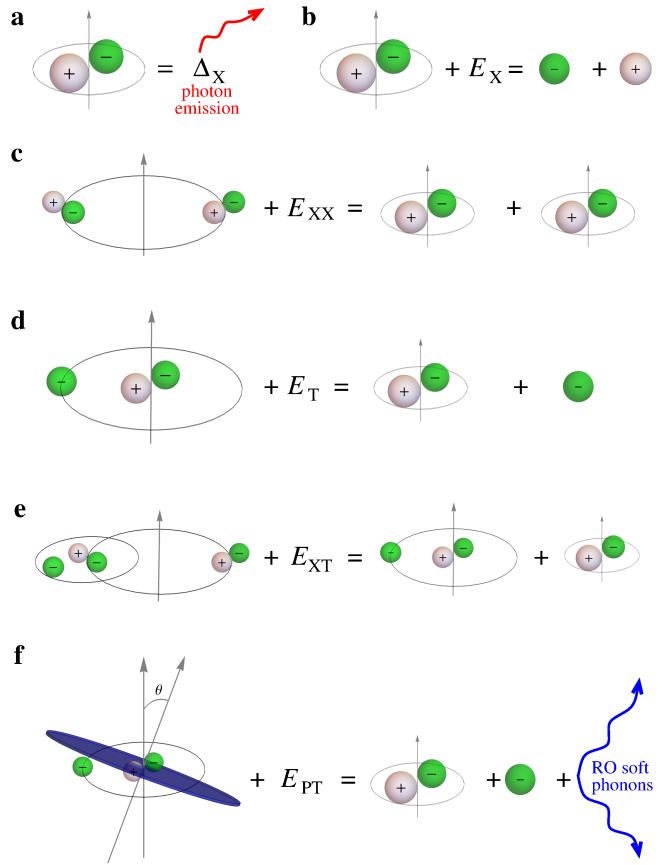}
  \caption{Real-space representation of quasiparticle decomposition with the binding energies indicated for each quasiparticle.
  {\bf a} Exciton recombination accompanied by emission of a photon with energy $\Delta_X$.
  {\bf b} Exciton ionization requires energy $E_X$. Interlayer excitons 
  may experience similar ionization processes with the electrons and holes remaining in their
  respective layers.
  {\bf c} To decompose a biexciton into two excitons, the energy $E_{XX}$ is required.
  {\bf d} A trion decays into an exciton and one free electron upon absorbing the trion binding energy $E_T$.
  {\bf e} A quinton with the binding energy $E_{XT}$ decays into a trion and an exciton.
  {\bf f} A polaronic trion possessing the binding energy $E_{PT}$ decays into an exciton, one free electron,
  and a few RO phonons emitted back into the substrate. 
  The angle, $\theta$, between trionic and phononic axes of rotation is indicated.
}
  \label{fig2}
\end{figure}

\subsection{Trions}

Trion can be seen as a bound state between an electron (or a hole)
and an exciton. Trions are also known as charged excitons,
which can be negative or positive depending on whether the excess charge is due to an electron or a hole.
The majority of conventional 2D semiconductors are n-doped so we focus on negatively charged trions here.
(The sign of the trion charge could, however, be altered
by electrostatic gating \cite{courtade2017charged}.)
The trionic binding energy is much lower than the excitonic one because
dipole-electron interactions are much weaker than electron-hole coupling.
The formal description starts from the coordinate transformation that allows the
relative and centre-of-mass exciton–electron motion to be separated. The trion Hamiltonian can be written as
\begin{equation}
H_T=\frac{\mathbf{P}_T^2}{2m_T} + \frac{\mathbf{p}_T^2}{2\mu_T} + V_T\left(\mathbf{r}_T\right),
\label{H_T}
\end{equation}
where $\mathbf{P}_T=-i\hbar \partial/\partial \mathbf{R}_T$, 
$\mathbf{p}_T=-i\hbar \partial/\partial \mathbf{r}_T$ are momenta operators
for the center-of-mass and relative electron-exciton motion, respectively,
and $V_T\left(\mathbf{r}_T\right)$ is the electron-exciton interaction
described by the dipole potential.
The relative and center-of-mass coordinates are defined as
$\mathbf{r}_T=\mathbf{R}_X - \mathbf{r}'_e$,
$\mathbf{R}_T=(m_X \mathbf{R}_X + m_e \mathbf{r}'_e)/m_T$,
where $\mathbf{r}'_e$ is the outer electron coordinate.
The total and reduced trion masses are $m_T=m_X+m_e$ and $\mu_T=m_X m_e/m_T$, respectively.
Solving the Schr\"odinger equation with the Hamiltonian $H_T$ one can find the theoretical trion binding energy $E_T$,
which is known to be of the order of 10 meV in 2D semiconductors \cite{kylanpaa2015binding,trions2017,falko2015trions}. This is the energy we need to decompose a trion into an electron and an exciton. 

Trions are in general less stable than excitons. 
Moreover, trionic decay may involve electron-hole recombination.
In contrast to the electron-hole recombination in excitons, where the energy is released as a photon, the trionic recombination results in a free electron absorbing excess energy.
The photon is then not emitted, which leads to a strong reduction in the PL quantum yield.
Hence, electron-exciton scattering limits the performance of PL devices based on 2D semiconductors. 
The limitation can be lifted by appropriate chemical or electrostatic doping,
which reduces excess electron concentration. Precluding trion formation and electron-exciton scattering
is the way to reach nearly 100\% PL efficiency \cite{Science2015javey,Science2019javey}.
In the opposite regime of high excess
electron concentrations the two-body Hamiltonian (\ref{H_T}) becomes inapplicable owing to the truly many-body interactions involved \cite{efimkin2017many}.
The resulting quasiparticle, known as a Fermi polaron, has been detected recently in 2D MoSe$_2$ \cite{sidler2017fermi}.

\subsection{Biexcitons}

Biexcitons are quasiparticles created by dipole-dipole interaction out of two excitons.
The dipole-dipole potential is weaker than electron-dipole interactions,
so the biexcitonic binding energy is lower than the trionic one \cite{PRB2017mostaani,trions2017,kezerashvili2017trion}.
Transforming coordinates in a similar way to before, we can write the biexciton Hamiltonian as
\begin{equation}
H_{XX}=\frac{\mathbf{P}_{XX}^2}{2m_{XX}} + \frac{\mathbf{p}_{XX}^2}{2\mu_{XX}} 
+ V_{XX}\left(\mathbf{r}_{XX}\right),
\label{H_XX}
\end{equation}
where $\mathbf{P}_{XX}=-i\hbar \partial/\partial \mathbf{R}_{XX}$, 
$\mathbf{p}_{XX}=-i\hbar \partial/\partial \mathbf{r}_{XX}$ are momenta operators
for the center-of-mass and relative exciton-exciton motion, respectively,
and $V_{XX}\left(\mathbf{r}_{XX}\right)$ is the exciton-exciton interaction
described by the dipole-dipole potential
$\propto r_X^2/r_{XX}^3$.
The relative and center-of-mass coordinates are respectively defined as
$\mathbf{r}_{XX}=\mathbf{R}_X - \mathbf{R}'_X$,
$\mathbf{R}_{XX}=(\mathbf{R}_X + \mathbf{R}'_X)/2$,
where $\mathbf{R}'_X$ is the second exciton center-of-mass coordinate, 
and $m_{XX}=2 m_X$, $\mu_{XX}=m_X/2$.

The dipole–dipole potential vanishes rapidly when the distance between the two excitons,
$\mathbf{r}_{XX}$ is much larger than the exciton size itself $\mathbf{r}_X$. This hinders biexciton formation, making it very challenging to detect the corresponding PL emission peak experimentally \cite{barbone2018complexes}.  Being a four-body process, the probability for exciton-exciton interactions increases quadratically with the excitonic density, and hence with the excitation power. 
Thus, parabolic dependence of PL intensity versus excitation power is a clear indication of biexcitonic effects involved.
Biexcitonic scattering, off each other or off defects or edges, may lead to recombination of one exciton and ionization of another, a phenomenon
known as exciton-exciton annihilation \cite{XXAoptical2014}. This is a non-radiative process that strongly reduces the PL quantum yield at high excitation power \cite{XXAparameters}. On the positive side, exciton-exciton annihilation produces high-energy electron-hole pairs that may be used in optically-excited transport to increase internal quantum efficiency \cite{XXAtransport2020}.

\subsection{Quintons}

A negatively charged quinton is a quasiparticle composed of three electrons and two holes. It can be described in two ways. First, a quinton can be seen as an electron bound to the quadruple core created by a biexciton \cite{PRB2017mostaani}.  Second, it can be seen as a bound state between a trion and an exciton. Note, however, that the most stable state should have the lowest potential energy. The quadruple potential is weaker than the dipole one; hence, the trion-exciton configuration should have a deeper energy state
than the electron–biexciton one. The quinton Hamiltonian can be written as
\begin{equation}
H_{XT}=\frac{\mathbf{P}_{XT}^2}{2m_{XT}} + \frac{\mathbf{p}_{XT}^2}{2\mu_{XT}} 
+ V_{XT}\left(\mathbf{r}_{XT}\right),
\label{H_XT}
\end{equation}
where $\mathbf{P}_{XT}=-i\hbar \partial/\partial \mathbf{R}_{XT}$, 
$\mathbf{p}_{XT}=-i\hbar \partial/\partial \mathbf{r}_{XT}$ are momenta operators for the center-of-mass and relative exciton-trion motion, respectively, and the exciton-trion interaction $V_{XT}\left(\mathbf{r}_{XT}\right)$ is given by potential between a point-like charge and a dipole representing a trion and an exciton, respectively.
The relative and center-of-mass coordinates are defined as
$\mathbf{r}_{XT}=\mathbf{R}'_X - \mathbf{R}_T$,
$\mathbf{R}_{XT}=(m_X\mathbf{R}'_X + m_T\mathbf{R}_T)/m_{XT}$ with 
the total and reduced quinton masses being $m_{XT}=m_X+m_T$ and $\mu_{XT}=m_X m_T/m_{XT}$, respectively.

Observation of quintons requires exceptional experimental conditions, including extreme purification of material components and cryogenic temperatures. The quasiparticle has been observed at 4K in 2D WSe$_2$ encapsulated between two layers of hexagonal boron nitride (h-BN) \cite{barbone2018complexes}.

\subsection{Polaronic trions}

Interactions possible in 2D semiconductors are not limited to the Coulombic ones forming excitons and exciton-like quasiparticles reviewed above. Electrons can couple with phonons by means of Fr\"ohlich interactions forming polarons \cite{froehlich1954}.
Although 2D semiconductors possess intrinsic
Fr\"ohlich interactions \cite{sohier2016two}, a much stronger polaronic coupling can be induced by an appropriate substrate,
for example, by strontium titanium oxide (STO)
\cite{devreese1996polarons}. The outer electron in a trion and soft rotational optical (RO) phonon modes in the STO substrate both rotate with the frequencies equivalent to 10 meV matching each other's symmetry and energy scale. The Fr\"ohlich coupling should, therefore, be efficient for trions, hence, creating a new quasiparticle
--- polaronic trion \cite{sarkar2019polaronic}.
The polaronic trion Hamiltonian can be written as
\begin{equation}
H_{PT}=\frac{\mathbf{P}_T^2}{2m_T} + \frac{\mathbf{p}_T^2}{2\mu_T} 
+ V_T\left(\mathbf{r}_T\right) + V_P\left(\mathbf{r}_T\right),
\label{H_PT}
\end{equation}
where $\mathbf{P}_T$, $\mathbf{p}_T$, and $V_T\left(\mathbf{r}_T\right)$ have been defined below Eq. (\ref{H_T}).
Here, we have introduced an additional interaction given by the polaronic term  $V_P\left(\mathbf{r}_T\right)$.
The corresponding Fourier transform can be derived 
following the Fr\"ohlich's recipe \cite{kittel1987quantum},
$V_P(q) = - i \hbar \omega \sqrt{2\pi\alpha r_\omega \cos\theta/q}$,
where $r_\omega = \sqrt{\hbar/2\mu_T \omega}$ is the interaction length, $\omega$ is the RO phonon frequency,
and $\alpha$ is standard Fr\"ohlich coupling constant \cite{froehlich1954}.
The striking difference between the standard 2D polaronic interaction and $V_P(q)$
is the $\sqrt{\cos\theta}$ pre-factor that occurs due to the special direction
singled out by the angular momentum of a RO phonon mode. The polaronic trion binding energy  therefore depends
on the crystallographic orientation of STO substrate with respect to the trion plane \cite{trushin2019evidence}.

\subsection{Interlayer excitons}

A pair of 2D semiconductors can be stacked together and excited optically.
Since the Coulomb interactions are screened weakly, they may extend across the interface providing 
coupling between electrons and holes excited in the neighbouring layers.
This forms interlayer excitons. The interlayer exciton Hamiltonian can be written as
\begin{equation}
H_{IX}=\frac{\mathbf{P}_X^2}{2m_X} + \frac{\mathbf{p}_X^2}{2\mu_X} 
+ V_{IX}\left(\mathbf{r}_X\right),
\label{H_IX}
\end{equation}
where $\mathbf{P}_X$ and $\mathbf{p}_X$ have been introduced below Eq. (\ref{H_X}),
and $V_{IX}\left(\mathbf{r}_X\right)$ is the interlayer electron-hole potential.
The interlayer interaction is weaker than the intralayer one because of the constant interlayer distance
$d$ always separating an electron and a hole.
The interlayer exciton binding energy $E_{IX}$ is therefore expected to be lower than $E_X$.
Note, however, that the interlayer bandgap $\Delta_I$ is strongly reduced because of the band offset between the
two layers, so that the corresponding PL emission energy given by $\Delta_{IX}=\Delta_I - E_{IX}$ turns out to be redshifted \cite{rivera2018interlayer} 
with respect to the intralayer excitonic peak $\Delta_X$.
The semiconducting layers can also be stacked with a twist that creates an additional {\em moir{\'e}} potential.
This potential splits the single interlayer excitonic peak into several resonances observed recently \cite{Nature2019evidence}.
The {\em moir{\'e}} excitons and exciton-like complexes are in the current focus of experimental   \cite{seyler2019signatures,jin2019observation,alexeev2019resonantly} and theoretical \cite{PRB2018complexes,interlayermoire2019} research.

The most important property of interlayer excitons is their extended recombination time due to the spatial separation of electrons and holes the excitons are composed of \cite{rivera2015observation}.
This makes it possible to study strongly correlated electron-hole states such as exciton condensation \cite{fogler2014high,PRBsuperfluidexciton2017,wang2019evidence} and electron-hole liquid \cite{arp2019electron-hole}. Recent papers on interlayer excitons have already been reviewed in Nat. Rev. Phys.\cite{tartakovskii2020excitons}.

\begin{figure}
  \includegraphics[width=\columnwidth]{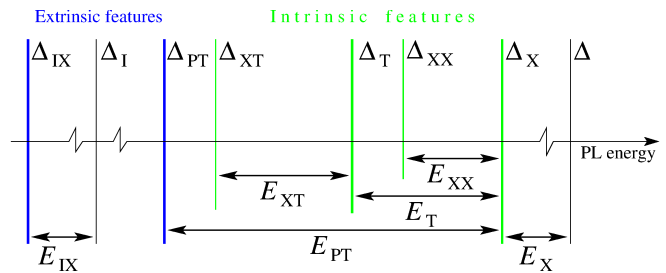}
  \caption{Relative positions of quasiparticle PL emission lines with the binding energies indicated. 
  (See Figs. \ref{fig1} and \ref{fig2} for notations.) Intrinsic features can be observed in isolated
2D semiconducting layers whereas extrinsic effects require either a special substrate or
another 2D layer nearby. Presentation inspired by \cite{PRB2017mostaani}.}
  \label{fig3}
\end{figure}

\subsection{Estimations and connections with experiments}
To conclude the section, we evaluate the accuracy of the Hamiltonians introduced above when compared with
data. We use the quasiparticle binding energies recently measured in high-quality 2D WSe$_2$ \cite{barbone2018complexes}:
$E_T=0.029$ eV, $E_{XX}=0.017$ eV, $E_{XT}=0.02$ eV.
The physical meaning of the energies is explained in Fig. \ref{fig2}. A typical PL spectrum is schematically shown in Fig. \ref{fig3}.
Excitonic binding energy is difficult to figure out from PL spectroscopy data, as the fundamental electronic bandgap does not produce any PL peak.
Using the band parameters for 2D WSe$_2$ $m_e=0.29m_0$, $m_h=0.34m_0$ \cite{PRB2013parameters} (here, $m_0$ is free electron mass),
and $\bar\epsilon=6.85$ (in-plane dielectric permittivity of h-BN \cite{NJP2018dielectric})
we estimate $E_X$ as $2\mu_X e^4/(\bar\epsilon^2 \hbar^2)\sim 0.18$ eV.

The quasiparticle size can then be estimated as a turning point of the relative quasiclassical motion in
the corresponding potential at a given binding energy. For excitons, this results in
$r_X= e^2/(\bar\epsilon E_X)$ being a little larger than 1 nm.
The biexciton size can be estimated assuming the dipole-dipole potential that suggests 
$r_{XX}\sim \sqrt[3]{e^2 r_X^2/(\bar\epsilon E_{XX})}$ to be a little more than 2 nm.
This is a reasonable value, as biexciton comprises two excitons with 1 nm size each.
Indeed, biexciton size cannot be much larger than $r_X$ because the dipole-dipole potential is not able to bind at such distances. Moreover, biexciton size smaller than $2r_X$ would be unreasonable because of electron-hole recombination annihilating the quasiparticle. Hence, the exciton and biexciton models are consistent.

Following the same approach, we quasiclassically estimate the trion size from 
the electron-dipole potential as $r_T\sim\sqrt{e^2 r_X/(\bar\epsilon E_T)}$
resulting in about 3 nm. The quinton size can be then estimated in the same way 
as $r_{XT}\sim\sqrt{e^2 r_X/(\bar\epsilon E_{XT})}$, which is about 4 nm. 
Hence, the quinton size turns out to be equal to the sum of the trion and exciton characteristic lengths. This again confirms the feasibility of the framework.

We do not consider spin and valley degrees of freedom
in our model Hamiltonians; however, 
the spin-valley locking intrinsic to 2D transition metal dichalcogenides may lead to the circularly polarized 
excitonic PL emission, with the chirality depending on which valley is excited \cite{mak2012control,cao2012valley,sallen2012robust,zeng2012valley}. Intervalley effects may also influence 
exciton linewidth \cite{glazov2014exciton,selig2016excitonic}
and lead to the splitting of single 
trionic PL peak into the intralayer and interlayer one \cite{plechinger2016trion}.

All the quasiparticles reviewed above can be seen as oscillators with the eigenfrequencies specific for each quasiparticle species.
The large oscillator strength of exciton-like quasiparticles in 2D semiconductors makes it possible
to strongly couple them with light in an optical cavity creating exciton-polaritons. From the theoretical point of view, an exciton-polariton can be described as two strongly coupled damped oscillators. The resulting eigenfrequency splits into two, and the splitting known as the Rabi frequency is observable in the emission spectra.
The 2D exciton-polaritons were initially detected in 2D MoS$_2$ \cite{liu2015strong} followed by
similar observations in 2D WS$_2$ \cite{flatten2016room,liu2017control,cuadra2018observation} and 2D MoSe$_2$ \cite{hu2017imaging}. 
Interestingly, polaritons inherit the valley selectivity (spin-valley locking) from the parent excitons \cite{chen2017valley,sun2017optical,dufferwiel2017valley}.
Theoretically, not only excitons but any other of the quasiparticles listed above may couple with light in a cavity to form polaritons.
This has already been experimentally confirmed for trions \cite{dhara2018anomalous}.
Hence, the polariton effect potentially doubles the number of exciton-like quasiparticle species.

Note, however, that to make quantitative predictions we should consider more realistic potentials
and use more exact solution techniques, such as GW-Bethe-Salpeter equation \cite{PRL2013louie} or quantum Monte Carlo \cite{PRB2017mostaani,PRB2018complexes}.

\begin{figure*}[ht!]
    \includegraphics[width = 6in]{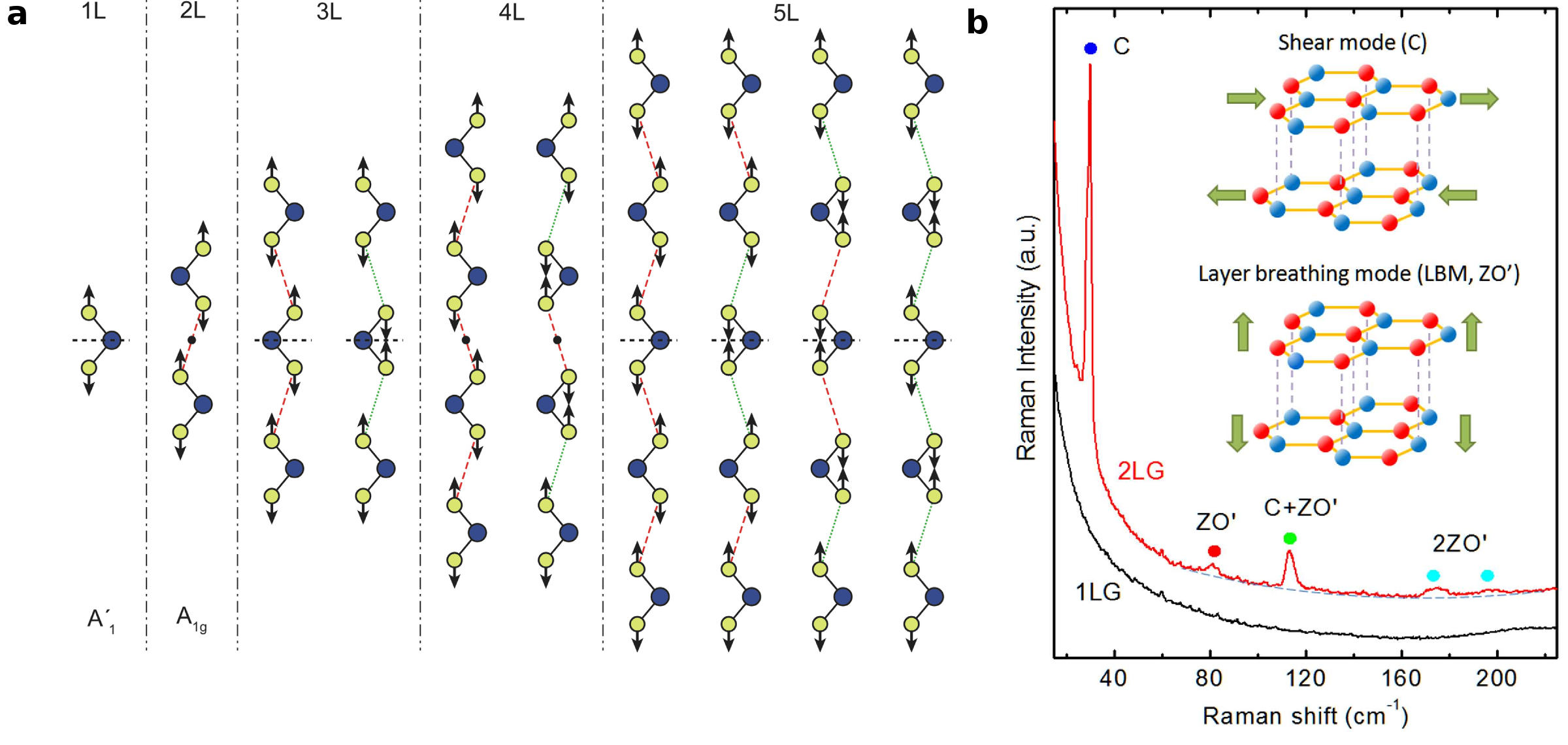}
    \caption{Emergence of new vibrational modes in multi-layer 2D materials.
    (a) Raman active out-of-plane vibrational modes in 1 to 5 layers of MoSe$_2$. Dashed red (green dotted) lines between the layers denote an increase (decrease) of the mode frequency compared to non-interacting layers. The horizontal dashed line and the dot indicate the mirror plane and inversion centre. (b)  Low-frequency Raman spectra from monolayer graphene (1LG, black line) and bilayer graphene (2LG, red line). The blue and red dots denote, respectively, the shear (C) mode and the layer breathing mode (LBM or ZO'). The green dot denotes the combination mode (C+ZO').  (a) is reproduced from Ref.\cite{tonndorf2013photoluminescence} with permission, OSA. (b) is reproduced from Ref.\cite{lui2014temperature} with permission, ACS. }
    \label{fig:phonons}
\end{figure*}

\section{Phonons}

In the 2D space, a crystal with $N$ atoms per unit cell has 2$N$ degrees of freedom, leading to the same number of vibrational modes (or phonons, in a quasiparticle language).
However, since they exist in the three-dimensional space, there are also out-of-plane vibrations, leading to a total of 3$N$ vibrational modes. These vibrations include, for example, flexural modes \cite{love1888xvi}, responsible for breaking the flatness of suspended monolayer materials at finite temperatures. In graphene, which has two atoms per unit cell, there are two flexural modes, the acoustical flexural mode, which corresponds at $k=0$ to the translation of the graphene along the direction normal to its plane, and the optical flexural mode, which corresponds to the out-of-phase oscillation of neighbouring atoms along the direction perpendicular to the graphene plane.
For long wavelengths, the acoustic flexural phonon mode (or just the ``flexural mode") has a minimal restoring torque, and a distinct parabolic dispersion\cite{jiang2015review,mariani2008flexural}.


In multi-layer systems, the number of atoms is multiplied and thus is the number of vibrational modes. Figure~\ref{fig:phonons}a illustrates this for the $A_1'$ mode of 2H-MoSe$_2$. Each monolayer mode gives rise to $N$ modes in the $N$-layer system (the figure shows Raman-active modes only). Because of inter-layer interactions, the frequency of the modes in the few-layer system is not the same as the monolayer system; in general, the bands split into different components, known as Davydov splitting. Additionally, few-layer systems may have different point group symmetry. Few-layer MoSe$_2$ with even number of layers has inversion symmetry, whereas monolayer and other systems with an odd number of layers do not. This symmetry difference results in selection rules turning on and off different vibrational modes depending on the number of layers.
The splitting of vibrational bands with the number of layers has been observed, for example in phosphorene\cite{favron-black-p} and MoTe$_2$\cite{song2016physical}.

The combination of acoustic (including flexural) vibrations in few-layer systems gives rise to shear modes and layer breathing modes when the layers move parallel to each other in different directions. Because inter-layer interactions are very weak compared to covalent intra-layer bonds, shear and inter-layer breathing modes are very low in energy - and therefore very difficult to observe directly in Raman spectroscopy, since they fall very close to the elastic scattering (Rayleigh) background. Their energy is also very sensitive to the layer number\cite{zhao2013interlayer,luo2015large,heinz-layer}
 
In graphene, breathing modes have been observed by means of a process involving two phonons in doubly resonant Raman spectroscopy\cite{heinz-layer} or employing the coupling between the phonon and the optical transitions\cite{shang2013observation}.
Shear modes have also been observed down to graphene bilayer using a modified setup\cite{tan2012shear}.
In few-layer phosphorene, the inter-layer breathing modes are Raman active and have been observed directly by Raman spectroscopy\cite{ling2015low,luo2015large}. Transition metal dichalcogenides also have Raman-active shear and inter-layer breathing modes that have been experimentally measured\cite{zhao2013interlayer,zhang2013raman,zhang2015phonon}.
Such success is driven by recent advances in Raman spectroscopy techniques\cite{liang2017low}.
Theoretically, such low-frequency modes are also difficult to model by density functional theory because of the small restoring forces involved compared with other crystal modes. However, they can be modelled by a simple linear chain model\cite{liang2017low,heinz-layer}.

We mentioned that flexural phonons have dispersion $\omega \propto k^2$\cite{review-graphene}, which has significant consequences.
Since the average of the population of phonons in mode $k$ is given by the Plank distribution,
\be\langle n(k)\rangle = [\exp(\beta D k^2)-1]^{-1},\label{eq:ph1}\ee
where $\beta$ is the $1/{K_bT}$ factor, the total number of phonons excited at temperature $T$ is given by:
\be n_k = \int_0^\infty g(k)\langle n(k)\rangle dk,\label{eq:ph2}\ee
where $g(k)$ is the density of states in 2D, $2\pi k$.
At long wavelengths, $\exp(\beta D k^2) \sim 1+ \beta D k^2$.
Thus, at any temperature $T>0$, the number of phonons diverges.
This divergence would have catastrophic consequences for the stability of 2D materials -- ruling out their existence altogether -- 
if not for the practical fact that 2D materials coexist with the surrounding 3D space, for example, the substrate
they are on, which damps oscillations.
As the number of layers increases, the dispersion of flexural modes also becomes increasingly linear in $k$\cite{jiang2015review}

However, free-standing monolayer membranes exist --- 
for example graphene drum-heads\cite{bunch2008impermeable} or other suspended graphene configurations.
The consequence of the phonon density of states $k$ dependence in 2D is that such suspended membranes are never really flat except at very low temperatures.
Mechanical vibrations are still damped even at resonance frequencies but can be excited using applied radio-frequency voltages\cite{garcia2008imaging}.
Graphene membranes are ideal for studying quantum motion. Because of their stiffness, the energies of the flexural modes are higher than cryogenic temperatures used in the experiments, ensuring the preservation of the quantum nature of these modes. The low mass of the membrane guarantees that the amplitude is high, allowing for easier detection of the zero-point motion\cite{song2014graphene,de2016tunable}.

Another significant consequence is the phonon contribution to limiting the carrier mobility in intrinsic 2D materials, e.g. MoS$_2$\cite{kaasbjerg2012phonon}.
However, in doped graphene, it has been found that the electron-phonon coupling to those phonons is strongly screened and negligible, and not detectable by transport measurements on doped graphene \cite{mauri}.

\begin{figure*}[ht!]
   \includegraphics[width=16cm]{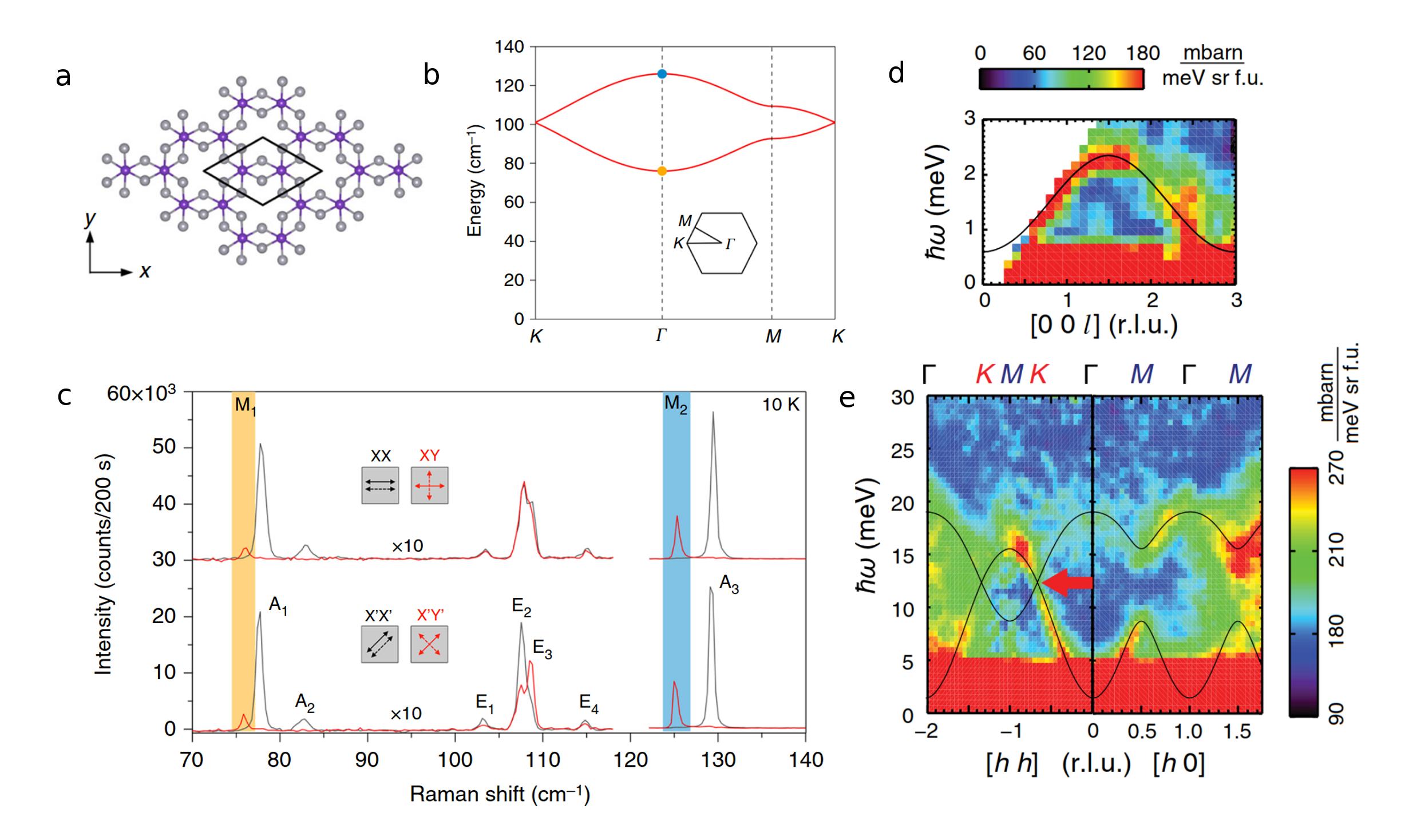}
  \caption{Magnon bandstructure in CrI$_3$, a honeycomb 2D material.
{\bf a}$|$ Atomic structure of a CrI$_3$ monolayer. Cr is represented in purple and I in gray.
{\bf b}$|$ Calculated magnon dispersion relation in monolayer CrI$_3$. The circles highlight zero-momentum modes.
{\bf c}$|$ Raman spectra of 13-layer CrI$_3$ at 10K. The magnon modes M$_1$ and M$_2$ are highlighted in
yellow and blue.
{\bf d}$|$ Magnon bandstructure as obtained from inelastic neutron scatering at 5K,
        integrated along the $l$ direction (reciprocal direction perpendicular to the $\Gamma-L-K$ plane)
        The superimposed solid line represents the Heisenberg-only theoretical model.
        The sample was a single-crystal platelet \cite{chen2018topological}.
        {\bf e}$|$ Low energy spin-wave mode along the $l$ direction. Minima are at $l=3n$ \cite{chen2018topological}.
Panels {\bf a-c} are reproduced with permission from REF.\onlinecite{jin2018raman}.
Panels {\bf d-e} are reproduced with permission from REF.\onlinecite{chen2018topological}.
\label{fig-m}
}
\end{figure*}

\section{Magnetic excitations}
\label{sec:Magnons}

The ability to manipulate the magnetic state of 2D crystals is of great interest for the design of future ultra-high-density magnetic storage devices, spintronics devices, such as spin valves, and topological quantum computing. Although magnetism in 2D crystals has remained elusive until very recently, the study of magnetic excitations has already been greatly advanced due to the ability to control, in 2D systems, the Fermi energy. In this section, we give a brief introduction to the controversy around the existence of ferromagnetism in 2D, magnons, topological excitations, and finally, spin density waves.

Magnons are collective excitations of the spins in magnetically ordered systems, i.e. ferromagnets and antiferromagnets. As the name indicates, they are quantised wave-like excitations, and therefore the term magnon is often used interchangeably with spin waves.

\subsection{Ferromagnetism}
The historical controversy around the existence of ferromagnetic order in 2D systems above zero temperature is intrinsically linked to the nature of 2D magnons. The semi-classical Ising model for a 2D system of spins perpendicular to the material plane finds the emergence of the long-range ordered ferromagnetic state below a critical temperature $T_c$. On the other hand, in its quantum mechanical counterpart, the Heisenberg model, where spins are allowed to point in any direction,
low energy magnons destroy such long-range order. More precisely, one can show that ferromagnetic or antiferromagnetic order is
impossible above 0~K for isotropic 2D systems with only short-range interactions.
This statement is known as the Mermin-Wagner theorem or Mermin-Wagner-Hohenberg theorem, proven rigorously by Mermin, Wagner and Hohenberg \cite{mermin1966absence,halperin2019hohenberg}. The reason for the breakdown of long-range order can be readily understood from the following statistical reasoning for an isotropic medium \cite{Vaz_2008}.
According to the 2D Heisenberg model, the dispersion of the lowest magnon band is quadratic\cite{coey2019permanent}:
\be\epsilon_{\bf k}=Dk^2.\ee
Thus, following similar reasoning to the one described for phonons in Eqs. (\ref{eq:ph1}--\ref{eq:ph2}),
we find that at any temperature $T>0$, the number of magnons diverges.

Nevertheless, numerous experimental observations in 2D systems,
starting with atomically thin epitaxial films of magnetic elemental systems (Co, Fe, Ni) \cite{cortie-advanced-review},
have left no doubt that magnetic order is possible in 2D. Early observations in a Fe monolayer film on Ni(001) have found that the magnetisation is perpendicular to the monolayer (eg.\cite{brien-PRB-52-15332}), contrary to what would be expected from the continuous theory for surfaces of 3D ferromagnets. It was then understood that finite-size effects, anisotropy, spin-orbit coupling, magnetic dipole interactions and other interactions would break the assumptions of Mermin-Wagner-Hohenberg. In reality, it turns out  that the finite size and anisotropy of real samples is too far removed from the conditions of the theorem.

Magnetic and ferromagnetic order has been observed in vdW materials and 2D sheets derived from vdW materials \cite{gong2019two}. Examples of 2D crystals include CrI$_3$ \cite{huang2017layer,jin2018raman}.
Cr$_2$Ge$_2$Te$_6$\cite{gong2017discovery} and FePS$_3$ \cite{lee2016ising}, with more predicted theoretically \cite{cortie-advanced-review,gong2019two}.

However, it should be noted that while such materials may be considered 2D from the point of view of structure or electronic interaction,
they still have significant exchange interaction across layers.
This is the case for example of CrI$_3$, where few-layer systems
with an odd number of layers display ferromagnetism in the direction perpendicular to the plane, but with even number of layers, the magnetisation along this direction is suppressed due to antiferromagnetic coupling between the layers \cite{huang2017layer}. Another example is Cr$_2$Ge$_2$Te$_6$, where $T_c$ vanishes in the monolayer limit,
a result that can be reproduced using an ideal 2D Heisenberg model with negligible single-ion anisotropy \cite{gong2017discovery}.

The quadratic magnon dispersion near $\Gamma$ (zero-momentum), as given by the isotropic Heisenberg model \cite{coey2019permanent}, is a good approximation near the Brillouin zone centre for most 2D crystals (see Fig.\ref{fig-m}a-d). Magnetic anisotropy, originating in the spin-orbit coupling, is a reason for the non-zero phonon energy at zero momentum (`spin-wave gap'), \cite{coey2019permanent,lado2017origin} shown in Fig.\ref{fig-m}b,d for CrI$_3$.

Magnon bandstructures for crystal lattices can be derived very similarly to phonon bandstructures. In honeycomb lattices, e.g., chromium trihalides CrX3 (X=F, Cl, Br and I), two of the magnon bands form Dirac cones near the $K$ point resembling the graphene electronic bandstructure \cite{Pershoguba-PRX-8-011010,fransson2016magnon}. In the case of CrI$_3$, these have been measured by inelastic neutron scattering (Fig.\ref{fig-m}e). However, there is a gap opening at the Dirac cone crossing due to Dzyaloshinskii-Moriya interaction \cite{chen2018topological}, a type of exchange interaction that favours spin canting.

\begin{figure*}
   \includegraphics[width=16cm]{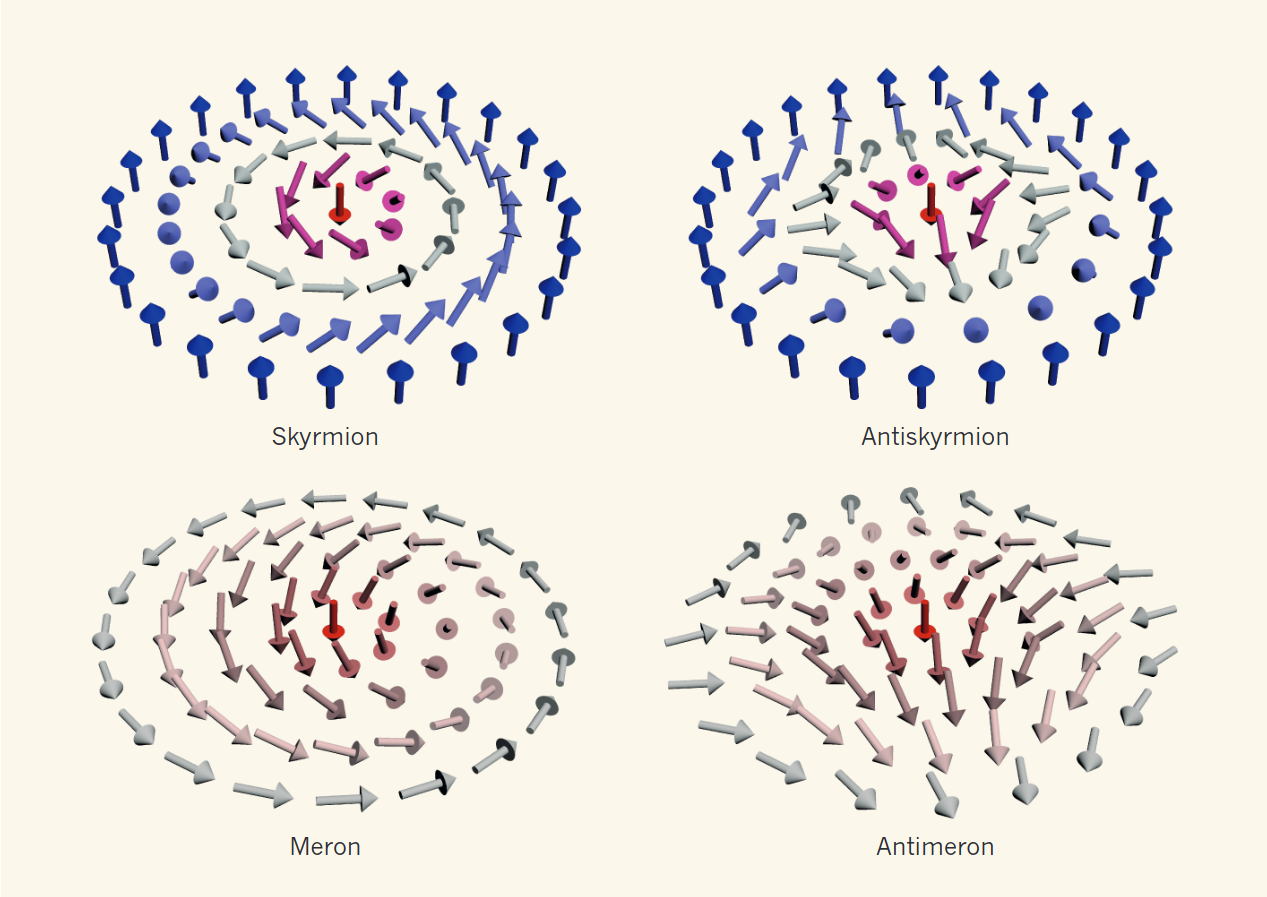}
  \caption{
          Chiral magnetic spin textures.
{\bf a}$|$ skyrmion (spiral-type), topological charge -1;
{\bf b}$|$ antiskyrmions, topological charge +1;
{\bf c}$|$ meron, topological charge $-1/2$;
{\bf d}$|$ antimeron, topological charge +$1/2$
Reproduced with permission from REF.\onlinecite{woo2018elusive}, Springer Nature Limited.
\label{fig-s}
}
\end{figure*}

\subsection{Topological excitations}
While in a 1D Ising model the simplest excitation is the flipping of one spin, in the corresponding Heisenberg model, this would correspond to a spin rotation on the plane perpendicular to the spin chain, or, equivalently, two domain walls.

In the 2D Heisenberg model, magnetic textures include vortices (in XY magnets), and skyrmions and merons in magnets with a non-collinear
spin ordering. Magnetic skyrmions are whirling configurations of magnetic momenta where the orientation of spins rotates progressively from the up direction at the edge to the down direction at the centre or vice versa (Fig.\ref{fig-s})\cite{skyrme1994non,breyskyrm,fert2013skyrmions}. More precisely, skyrmions (and anti-skyrmions) are characterised by an integer topological charge,
\be Q=\frac{1}{4\pi}\int d^2r(\partial_x{\bf m}\times\partial_y{\bf m})\cdot {\bf m},\ee
where ${\bf m}$ is the order parameter (eg. a unit vector along the direction of magnetisation).
$Q$ describes how many times the magnetic moments wrap around a unit sphere upon application of stereographic projection \cite{kovalev-frontiers-physics-27-sep-2018}.
Merons have been described as half-skyrmions and have $Q=-1/2$ ($Q=1/2$ for antimerons) \cite{yu2018transformation}.
While the lattice magnons considered in the previous section can be seen as quantised travelling waves and thus 
carry lattice momentum, skyrmions can carry angular momentum.

In 2D materials, skyrmions can be conveniently detected and manipulated using surface microscopy techniques.
Some of the material characteristics to look for in 2D materials that may support skyrmions are a strong spin-orbit coupling, and additionally a Dzyaloshinskii–Moriya interaction, which favours spin canting and seems to often play an important role in stabilising skyrmions \cite{banerjee2014enhanced}.

A topological phase transition, driven by the un-binding of vortex and anti-vortex pairs with the increase of temperature
in XY ferromagnets, was proposed by Berezinskii and by Kosterlitz and  Thouless \cite{sorokin2019critical}. Spin vortex phase transitions are still a matter of on-going research.

\subsection{Itinerant ferromagnetism}
Itinerant ferromagnetism in metals originates with the conduction (itinerant) electrons with an extended wavefunction.
In a one-electron picture, this happens when the number of filled states for spin-up electrons and spin-down electrons is different (the difference can even be fractional).
For example, SnO, GaS, and GaSe have electron bands with Mexican hat dispersion, which have been shown to
give rise to a ferromagnetic state if the Fermi level is chosen to lie in the nearly-flat region \cite{SnO}.
In twisted bilayer graphene  {\em moir{\'e}} superlattices, close to a critical twist angle when the density of states at the Fermi energy becomes very larges, ferromagnetic states emerge due to increasing electron-electron interaction\cite{sharpe2019emergent}.
Nevertheless, no ferromagnets of this type with Curie temperature close to room temperature are known to date.

\subsection{Spin density waves}
Spin density waves are broken-symmetry ground states of metals, with a periodic modulation of the charge density that is not necessarily commensurate with the lattice period.
Different from the spin waves (or magnons) considered in the preceding session,
it is not the spin orientation that varies but the spin density.
Charge density waves can have low-lying spin, charge and phase excitations.
The spin density can vary in space
in metallic systems where there is a large density of states near the
Fermi energy, particularly if there is nesting of the Fermi surface \cite{schlottmann-AIPadvances-6-055701}.
For this reason, spin density waves, charge density waves and superconductivity often occur in
the same 2D systems and their interplay is still an object of research.

\section{Outlook}
In the decade and a half since the discovery of graphene, 2D materials have become an integral part of physics
and materials science research has been driven by rapid scientific progress. This increased understanding has opened new
frontiers in the study of 2D systems.

In the field of plasmonics, one of the most critical challenges has to do with augmenting the plasmon
propagation length, which can be achieved by careful sample engineering. This improvement would allow the use of plasmons as signal carriers in plasmonic devices.
Improved sample preparation techniques would also expand the plasmonic field from graphene to include materials such as phosphorene, predicted to host anisotropic plasmons.
Finally, ultraclean samples make it possible to probe the hydrodynamic regime, where Lindhard treatment of electron–electron interaction becomes insufficient.
The theoretical formalism of these hydrodynamic plasmons is currently an active area of research.

In the excitonic field, there are two main directions whose intertwining may result in non-trivial physics and
applications. The directions are moiré excitons in twisted 2D semiconducting bilayers, and excitonic condensation in 2D homo- and heterostructures. Facilitating interactions by means
of moir\'e excitons and making use of the increased interlayer excitonic lifetime in excitonic condensation may produce synergistic effects yet to be discovered.

Phonon dispersion simulations in 2D materials are still very challenging, making it difficult to assess phonon-limited mobility and phonon-mediated heat transport.
It will be necessary in future to quantify such effects in 2D materials on substrates and in heterostructures.

Since 2D magnetic materials have been realized, it is possible to study domain wall motion and topological excitations. There is ample room to find new physics; 
the barrier to applications seems to be the small Curie temperature of the materials that are known so far.

In general, the effect of electron–electron correlation and its impact in all these crystal excitations is still not well understood. New experiments on moir\'e lattices
of twisted layers may be a way to vary correlation and understand its effects. The reason that spin density waves, magnetism and superconductivity coexist in 
some materials and the definition of their phase diagram is also still an open problem.

Lastly, it can be said that any 2D material is its surface; this provides a window to observe in real space any type of collective excitation in an unprecedented way.
Paraphrasing Isaac Newton, at this point in time, we are only children playing on a beach, while vast oceans of truth lie undiscovered before us.

\section{Acknowledgements}
This work was supported by the National Research Foundation, Prime Minister Office, Singapore,
under its Medium Sized Centre Programme and CRP
award ``Novel 2D materials with tailored properties: beyond graphene" (Grant number R-144-000-295-281). A. R. acknowledges support by Yale-NUS College (through Grant No. R-607-265-380-121). M.T. is supported by the Director's
Senior Research Fellowship from the Centre for Advanced 2D
Materials.

\bibliographystyle{naturemag}
\bibliography{2D.bib}

\end{document}